\documentclass[twocolumn,aps,prstab,showpacs,groupedaddress,10pt]{revtex4-1}

\usepackage{graphicx}
\usepackage{multirow}

\begin{document}

\title{Particle production and energy deposition studies for the 
neutrino factory target station}

\author{John J. Back}
\email[]{J.J.Back@warwick.ac.uk}
\affiliation{University of Warwick, Coventry, CV4 7AL, United Kingdom}
\author{Chris Densham}
\affiliation{STFC Rutherford Appleton Laboratory, Didcot,
OX11 0QX, United Kingdom}
\author{Rob Edgecock}
\affiliation{STFC Rutherford Appleton Laboratory, Didcot,
OX11 0QX, United Kingdom, \\ 
and Huddersfield University, Huddersfield, HD1 3DH, United Kingdom}
\author{Gersende Prior}
\affiliation{European Organization for Nuclear Research, CERN CH-1211,
Gen\`{e}ve 23, Switzerland}

\date{February 1, 2013}

\begin{abstract}
We present FLUKA and MARS simulation studies of the pion production and 
energy deposition in the Neutrino Factory baseline target station, which
consists of a 4\,MW proton beam interacting with a liquid mercury jet target
within a 20\,T solenoidal magnetic field. We show that a substantial
increase in the shielding is needed to protect the superconducting coils
from too much energy deposition. Investigations reveal that it is 
possible to reduce the magnetic field in the solenoid capture system 
without adversely affecting the pion production efficiency.
We show estimates of the amount of concrete shielding
that will be required to protect the environment from the high radiation
doses generated by the target station facility. We also present yield and energy 
deposition results for alternative targets: gallium liquid jet, 
tungsten powder jet, and solid tungsten bars.

\end{abstract}


\maketitle

\section{Introduction\label{sec:Intro}}

The current baseline option for the Neutrino Factory~\cite{ref:NuFact} is to use
a 4\,MW proton beam interacting with a free-flowing mercury jet
to create an intense muon beam~\cite{ref:IDSNF}.
The MERIT experiment has shown a proof-of-principle demonstration of
a high intensity liquid mercury jet target~\cite{ref:MERIT}.
The interaction of the bunched proton beam (rms bunch length equal to 3\,ns)
with the mercury jet creates low-energy pions that are captured by the high field 
($\sim20$\,T) solenoid and transported through a decay channel. Muons
resulting from the decay of these pions pass through a cooling section and 
circulate around a storage ring until they decay to neutrinos.

In this paper, we present a series of simulation studies, using
the FLUKA~\cite{ref:FLUKACode} and MARS~\cite{ref:MARSCode} computer packages, on
particle production and energy deposition (radiation dose) calculations.
We first show the simulation results for the so-called Study 2 geometry configuration, 
and then describe how the geometry needs to be modified to address  
various safety issues. We also compare useful muon yields and energy deposition
doses for different target material alternatives. Finally, we present
a study on the concrete shielding requirements that will be necessary to protect
the environment from the high radiation doses emanating from the target station.

\section{Simulation Parameters\label{sec:SimParams}}

\begin{figure}[!hbt]
\includegraphics[width=\columnwidth]{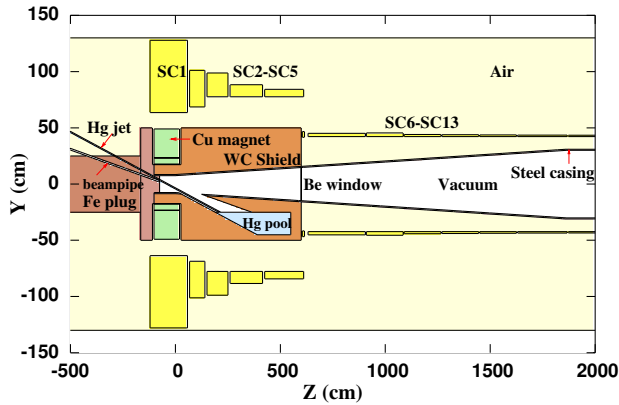}
\caption{Schematic of the Study 2 geometry of the Neutrino Factory target system.
The shielding is comprised of tungsten carbide (80\%) with water cooling (20\%).
The superconducting magnets are labelled SC$n$, where $n$ = 1 to 13.
\label{fig:Study2Geometry}}
\end{figure}

As a starting point, the Neutrino Factory mercury jet target station
geometry is based on the Study 2 configuration~\cite{ref:Study2}, as shown
in Fig.~\ref{fig:Study2Geometry}, with the appropriate 20\,T field map
based on the dimensions and currents in the normal-conducting copper ($\sim 6$\,T)
and superconducting ($\sim 14$\,T) coils.
Variations to the geometry are made to reduce the energy deposition
in the superconducting magnets, as well as incorporating more engineering
considerations to the overall target station design.
In all simulations, the proton beam has a transverse Gaussian profile with a root mean 
square radius of 1.2\,mm. The kinetic energy of the proton beam is nominally set to 8\,GeV, 
but is varied when finding the optimal number of useful muons from the target.
The mercury jet is modeled as a simple cylinder with a radius of 4\,mm, tilted
at approximately 100\,mrad to the magnetic $z$ axis.
We also investigate the yields and energy deposition for alternative targets, also
tilted at 100\,mrad to the $z$ axis:
liquid gallium jet, powder tungsten jet (50\% density), and solid tungsten bars. The 
first two alternatives are also modeled as cylinders with a radius of 4\,mm, while 
the solid target is modeled as a 20\,cm long, 2\,cm diameter cylinder
with a density of 19.25\,g/cc.
The angle between the target and the proton beam at their intersection
($z = -37.5$\,cm) varies between 20 and 30\,mrad, depending
on the initial kinetic energy, in order to optimize pion 
production~\cite{ref:DingStudy}.

\section{Study 2 Geometry\label{sec:Study2}}

\subsection{Muon yields\label{sec:Study2Yields}}

One important figure of merit concerning the performance of the target is the total number
of muons that pass through the Neutrino Factory cooling channel. This is calculated
by first counting the number of pions, kaons, and muons that are directly produced
by the mercury jet-proton beam interaction. These secondary particles are then tracked
through the solenoidal target decay channel up to $z=50$\,m. The ICOOL simulation 
package~\cite{ref:ICOOL} is used to 
find what fraction of these secondary particles end up as muons within
the accelerator acceptance (30\,mm transversely and 150\,mm longitudinally, with
$z$-momenta between 100 and 300\,MeV/c).

\begin{figure}[!hbt]
\includegraphics[width=0.775\columnwidth,angle=-90]{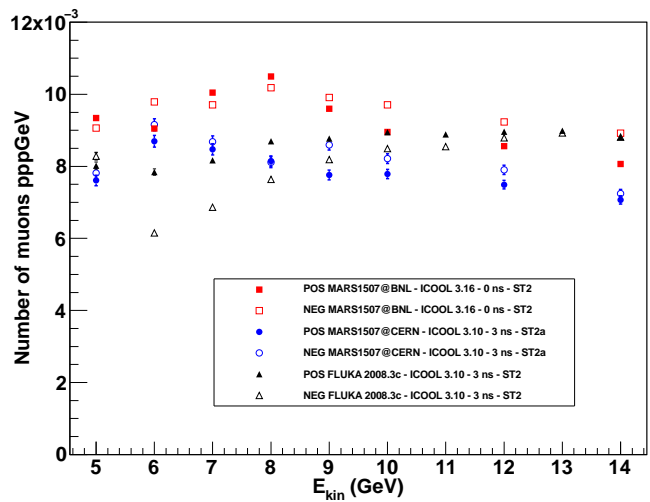}
\caption{The accepted muon yield per proton per GeV as a function of proton beam kinetic
energy. The different points show what effect different simulation code versions
have on the yield calculation results. ST2a and ST2 denote
different $\underline{B}$ field tapering parameters. For ST2a (ST2), $|\underline{B}|$ 
adiabatically decreases from 20 to 1.75\,T (1.25\,T) over a length of 12\,m (18\,m).
\label{fig:Study2Yield}}
\end{figure}

Figure~\ref{fig:Study2Yield} shows the expected muon yields
for the Study 2 (ST2) and Study 2a (ST2a) target configurations, as a 
function of the kinetic energy of the incoming proton beam. 
These yields are normalized to the number of protons on target as 
well as the proton beam kinetic energy. In general, the results are in 
agreement with the model-independent conclusions of Ref.~\cite{ref:Strait} that ``the 
dependence of the muon yield on proton beam energy at constant beam 
power is relatively flat, and any energy between 4 and 11\,GeV has a 
yield that is within 10\% of the maximum at 7\,GeV.'' There is some 
variation in the calculated muon yield, depending on what version of 
the simulation codes are used, as well as on the assumptions made 
about the proton beam bunch spacing (0 or 3\,ns). For the former, this 
can only be improved by updates to the simulation codes.

\subsection{Energy deposition\label{sec:Study2PowerDep}}

\begin{figure}[!hbt]
\includegraphics[width=\columnwidth]{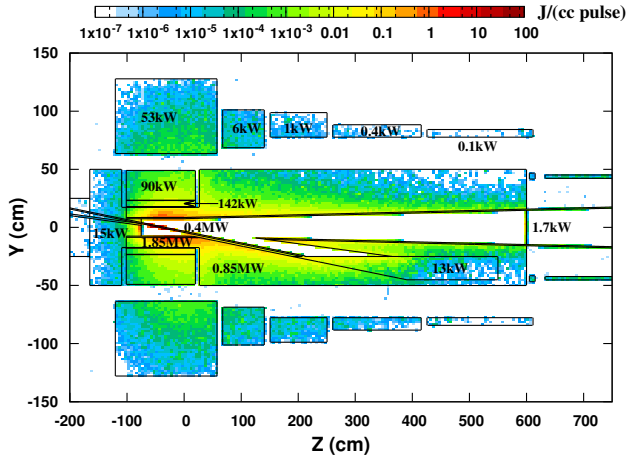}
\caption{Distribution of the deposited energy density (J/cc) per beam pulse
(50\,Hz repetition rate) in the Study 2 target system. Also shown is the estimated
deposited power within various regions.
\label{fig:Study2PowerDep}}
\end{figure}

Figure~\ref{fig:Study2PowerDep} shows the distribution of energy deposition
in the Study 2 target station using the FLUKA simulation code, which agrees
rather well with those results obtained using MARS~\cite{ref:MARSPower}. 
Also shown is the total power dissipated in the first few superconducting coils. 
The values in SC1--SC3 are unacceptably high and certainly have to be reduced by 
increasing the shielding thickness~\cite{ref:JinstEDep}.
Previous studies have shown that the forces between the superconducting
coils are very high~\cite{ref:MagForces}, and in the present geometry there is no space for
support structures between individual coils, which needs to be addressed.
Another issue is that the normal conducting magnets also experience very high 
levels of power dissipation (142 and 90\,kW). This problem is avoided by removing them 
completely and replacing the volume with more shielding. This means that the magnetic
field in the beam-jet interaction region is decreased from 20\,T down to 15\,T.
Also note that the power deposition in the mercury pool is quite low (13\,kW).
This means that the volume of the pool reservoir needs to be increased
in order for it to act as an effective beam dump for protons that do not interact 
with the mercury jet. In the next section 
we explore what effect these necessary geometry changes have on both the distribution 
of the energy deposition and on the accepted muon yield efficiencies.

\section{Improved Shielding Configuration\label{sec:IDS120j}}

Figure~\ref{fig:IDS120jGeometry} shows the new target station geometry configuration
that has increased shielding (approximately doubled in thickness laterally) to 
protect the superconducting (SC) coils, with the normal conducting magnets removed, as 
well as a larger mercury pool reservoir (88\,cm $<z<$ 367\,cm, $y<-15$\,cm, $r<45$\,cm). 
Table~\ref{tab:IDS120jSCCoils} provides the geometrical and current density
parameters of the SC coils. 
The SC coils are arranged in triplets, each corresponding to one cryostat module.
Gaps are introduced between neighbouring coils to provide space for the cryostat
modules and supporting structures to hold them in place
(the cooling components and their internal shielding are not included in the simulation).
Each cryostat is protected by large shielding volumes, made from tungsten beads (60\%)
with an assumed low-grade density of 15.8\,g/cc, and fast-flowing helium gas (40\%), which 
replaces water as the cooling agent. This change is motivated
by the concern that any creation of bubbles in the water will affect its circulation,
giving nonuniform cooling of the tungsten beads, as well as the 
problem of corrosion in a high radiation environment~\cite{ref:WCorrosion}.
Each shielding section, separated by 20\,cm gaps, is surrounded by stainless steel 
container vessels (each with a thickness between 2 and 10\,cm) that must support
the $\sim$200\,tonne weight while limiting stresses and deformations
to acceptable values~\cite{ref:CoilSupport}.
{
\begin{table}[hbt!]
\caption{Parameters defining the superconducting
(SC) coils for the new shielding geometry shown in Fig.~\ref{fig:IDS120jGeometry}: 
$z_0$ is the initial $z$ position, $\Delta z$ specifies the length along $z$, $r_1$ is 
the inner radius, $\Delta r$ is the radial thickness, while $I$ is the average current density.
The last column specifies the coil materials used in the FLUKA simulation, where SCon is the
compound Cu (54\%), Nb (24\%), Ti (12\%), and kapton (10\%). The mass densities for
Nb$_3$Sn and SCon are 6.8\,g/cc and 7.0\,g/cc, respectively.
\label{tab:IDS120jSCCoils}}
\begin{ruledtabular}
\begin{tabular}{lllllll}
Coil  & $z_0$      & $\Delta z$   & $r_1$      & $\Delta r$     & $I$            & Material \\
      & (cm)       & (cm)         & (cm)       & (cm)           & (A/mm$^2$)     & \\
\hline
SC1   & $-$240.5     & 355.0        & 120.0      & 75.8           &  19.3          & Nb$_3$Sn \\
SC2   &  114.5     & 72.7         & 120.0      & 64.3           &  22.0          & Nb$_3$Sn \\
SC3   &  273.6     & 48.1         & 120.0      & 75.8           &  26.7          & Nb$_3$Sn \\
SC4   &  459.0     & 21.3         & 90.0       & 57.6           &  33.8          & Nb$_3$Sn \\
SC5   &  534.6     & 319.7        & 90.0       &  4.7           &  40.9          & Nb$_3$Sn \\
SC6   &  929.8     & 11.2         & 90.0       & 50.6           &  41.9          & Nb$_3$Sn \\
SC7   &  1036.0    & 10.7         & 70.0       & 20.0           &  45.0          & SCon \\
SC8   &  1081.7    & 339.5        & 70.0       &  2.5           &  46.7          & SCon \\
SC9   &  1453.0    & 11.0         & 70.0       & 20.0           &  46.3          & SCon \\
SC10  &  1534.7    & 10.7         & 70.0       & 20.0           &  45.8          & SCon \\
SC11  &  1575.8    & 348.3        & 70.0       &  2.5           &  47.7          & SCon \\
SC12  &  1960.3    & 11.0         & 70.0       & 20.0           &  45.8          & SCon \\
\end{tabular}
\end{ruledtabular}
\end{table}

The beam pipe is modeled as a 2\,cm-thick stainless steel tapered volume,
which defines the inner bore of the decay vacuum region. The beam pipe section
just above the mercury pool surface is removed between $z=88$ and 200\,cm
to allow the mercury jet (and noninteracting protons) to enter the pool unhindered.
The beryllium window, which separates the 
jet-beam interaction region from the rest of the decay channel vacuum, is moved further
upstream to coincide with the edge of the first cryostat module. It consists 
of a 0.5\,cm gap containing rapidly flowing He gas for cooling
sandwiched between two 0.5\,cm Be layers.
The iron yoke plug has been removed to allow for space for the mercury jet nozzle 
injection and return flow system (which is ignored in the simulation). 
The mercury jet is still modeled as a simple cylinder of radius 4\,mm, tilted at 
approximately 100\,mrad with respect to the magnetic $z$ axis.

\begin{figure}[!hbt]
\includegraphics[width=\columnwidth]{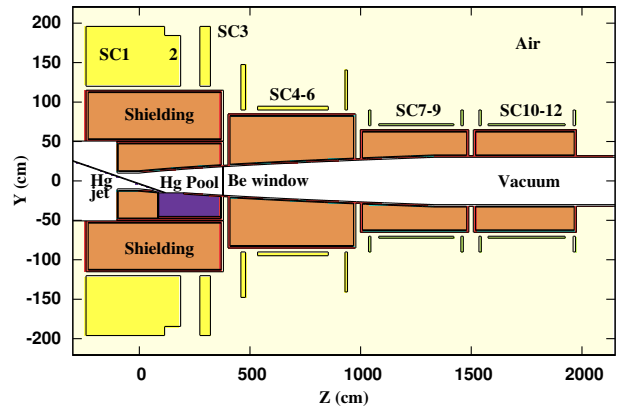}
\caption{Schematic of the updated target geometry configuration with
increased shielding. The superconducting magnets are labelled SC$n$, where 
$n$ = 1 to 12.
\label{fig:IDS120jGeometry}}
\end{figure}

\subsection{Pion and muon yields\label{sec:IDS120jYields}}

The removal of the normal conducting magnets implies 
that the peak magnetic field in the jet-beam interaction region is 
decreased from 20 to 15\,T. To ensure that secondary pions from the target 
maintain their magnetic rigidity, the inner bore radius of the beam pipe at
this location is increased from 7.5 to 10\,cm.
Figure~\ref{fig:IDS120jBField} shows a comparison of the axial
magnetic field profile between the original Study 2 geometry
and the new increased shielding geometry. The latter has a much
wider magnetic field profile, which helps to capture much more secondary
charged particles from the target. This can be clearly seen in
Fig.~\ref{fig:IDS120jYield}, which shows the improved charged averaged yield
of useful pions and muons from the target, normalized to the total
number of protons on target and the initial beam kinetic energy.
Using the FLUKA simulation package, these yields are calculated by 
finding the number of pions and muons (of both signs) that pass a 
transverse plane 50\,m downstream from the beam-jet interaction region,
within the decay channel aperture that has a bore radius of 30\,cm.
To obtain a figure of merit for the muon yield for the Neutrino Factory, we 
require that these particles have kinetic energies between 40 and 
180\,MeV~\cite{ref:DingStudy}.

In addition to the baseline mercury jet case, the yields
from alternative target materials are also shown in Fig.~\ref{fig:IDS120jYield},
using the simulation parameters outlined in Sec.~\ref{sec:SimParams}.
Between 5 and 15\,GeV, there is a two-peak structure in the normalised
yield distributions, owing to the transition between different hadronic
models used at low- and high-energies in the FLUKA simulation code.
Above 5\,GeV, the mercury jet provides the best yields, although the 
performance of the solid tungsten target matches this very closely. 
Note that we do not see a dramatic reduction in the yield for the tungsten powder jet,
which is assumed to have an effective density of 50\% of solid tungsten. The reason
for this is that, even though less protons will interact with the powder target, the
amount of reabsorption of secondary particles inside the target will also be lower, 
giving an overall figure of merit comparable to either the mercury jet or solid tungsten 
case. At low beam kinetic energy (below 5\,GeV), there is an indication that the
liquid gallium target gives the best yield compared to the other materials.
Further analysis is required to optimize the yields for this improved shielding geometry.
\begin{figure}[!htb]
\includegraphics[width=\columnwidth]{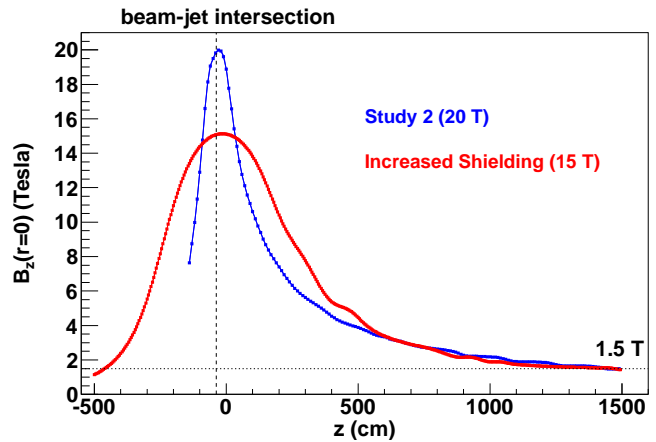}
\caption{Axial magnetic field profile distributions for the Study 2 and new
increased shielding geometries.
\label{fig:IDS120jBField}}
\end{figure}
\begin{figure}[!hbt]
\includegraphics[width=\columnwidth]{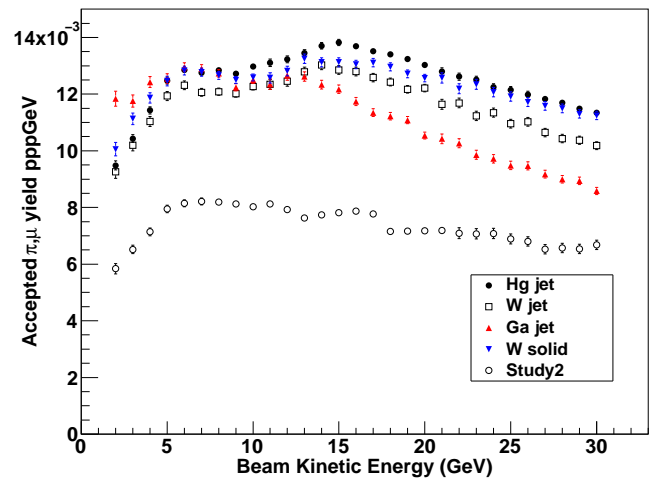}
\caption{The charged-averaged accepted pion and muon yield per proton per GeV for various
targets in the new increased shielding geometry. Also shown are the equivalent yields
for the mercury jet target in the Study 2 geometry.
\label{fig:IDS120jYield}}
\end{figure}

\subsection{Energy deposition\label{sec:IDS120jPowerDep}}

{
\begin{table*}[htb!]
\caption{Power deposition in various regions of the increased shielding
geometry configuration for the mercury jet, as well as for
alternative target materials. For the solid tungsten bars, the pool
reservoir is replaced by more shielding.\label{tab:IDS120jPower}}
\begin{ruledtabular}
\begin{tabular}{lllll}
& \multicolumn{4}{c}{Deposited power (kW)}\\
Region & Hg jet & Ga jet & W powder jet & W solid \\
\hline
SC coils 1--12 & $0.57 \pm 0.05$ & $0.67 \pm 0.06$ & $0.62 \pm 0.06$ & $0.55 \pm 0.06$ \\
Lower shielding SC 1--3 ($r < 50$\,cm, $z < 83$\,cm) & $1284.4 \pm 8.3$ & $1034.9 \pm 8.1$ & $1154.3 \pm 9.0$ & $1282.6 \pm 7.1$ \\
Lower shielding SC 1--3 ($r < 50$\,cm, $z > 83$\,cm) & $234.2 \pm 3.5$ & $318.5 \pm 3.6$ & $284.8 \pm 4.0$ & $348.2 \pm 6.2$ \\
Upper shielding SC 1--3 ($r > 50$\,cm) & $58.3 \pm 0.8$ & $82.6 \pm 1.1$ & $75.8 \pm 1.0$ & $41.9 \pm 0.5$ \\
Shielding for SC 4--6 & $38.0 \pm 1.7$ & $45.2 \pm 2.0$ & $38.8 \pm 1.9$ & $26.1 \pm 1.2$ \\
Shielding for SC 7--9 & $11.0 \pm 0.7$ & $11.9 \pm 0.8$ & $10.8 \pm 0.8$ & $8.1 \pm 0.6$ \\
Shielding for SC 10--12 & $7.4 \pm 0.7$ & $8.1 \pm 0.7$ & $7.1 \pm 0.8$ & $5.0 \pm 0.5$ \\
Beam pipe up to $z = 0$\,cm & $352.3 \pm 2.9$ & $230.8 \pm 2.3$ & $303.2 \pm 3.3$ & $303.0 \pm 1.9$ \\
Beam pipe from $z = 0$\,cm to end of taper & $397.6 \pm 3.6$ & $499.1 \pm 4.3$ & $428.7 \pm 3.8$ & $338.8 \pm 4.1$\\
Beam pipe from end of taper & $21.7 \pm 0.9$ & $24.0 \pm 1.0$ & $21.1 \pm 1.0$ & $14.6 \pm 0.8$ \\
Lower shielding vessel for SC 1--3 ($r < 50$\,cm) & $7.6 \pm 0.2$ & $12.5 \pm 0.3$ & $9.7 \pm 0.2$ & $5.7 \pm 0.2$ \\
Upper shielding vessel for SC 1--3 ($r > 50$\,cm) & $6.0 \pm 0.1$ & $9.3 \pm 0.2$ & $7.8 \pm 0.2$ & $4.3 \pm 0.1$ \\
Shielding vessel for SC 4--6 & $3.5 \pm 0.3$ & $4.5 \pm 0.3$ & $3.6 \pm 0.3$ & $2.4 \pm 0.2$ \\
Shielding vessel for SC 7--9 & $0.8 \pm 0.1$ & $0.8 \pm 0.1$ & $0.7 \pm 0.1$ & $0.6 \pm 0.1$ \\
Shielding vessel for SC 10--12 & $0.5 \pm 0.1$ & $0.6 \pm 0.1$ & $0.5 \pm 0.1$ & $0.3 \pm 0.1$ \\
Pool reservoir container & $10.5 \pm 0.3$ & $17.1 \pm 0.4$ & $14.0 \pm 0.4$ &  \\
Pool reservoir & $460.8 \pm 9.7$ & $814.1 \pm 10.6$ & $655.3 \pm 11.5$ &  \\
Jet/target & $416.8 \pm 2.4$ & $167.3 \pm 1.0$ & $298.7 \pm 2.3$ & $1018.5 \pm 5.2$ \\
Be window & $8.9 \pm 0.1$ & $6.3 \pm 0.1$ & $8.4 \pm 0.1$ & $5.1 \pm 0.1$ \\
Total & $3320.7 \pm 14.4$ & $3288.1 \pm 15.0$ & $3323.8 \pm 16.3$ & $3405.6 \pm 11.8$ \\
\end{tabular}
\end{ruledtabular}
\end{table*}
}

We have seen that the change to the target station geometry has had a positive
effect to the useful muon yield. The situation regarding the energy deposition
for the superconducting coils is also improved. Figure~\ref{fig:IDS120jPowerDep}
shows the energy deposition in the new target station geometry, obtained
using the FLUKA simulation code, which shows that the increase to the 
shielding has dramatically reduced the power dissipation in the SC coils.
\begin{figure}[!hbt]
\includegraphics[width=\columnwidth]{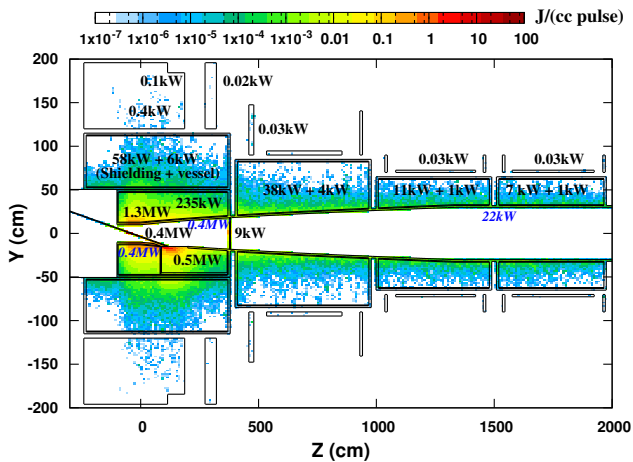}
\caption{Distribution of the deposited energy density (J/cc) per beam
pulse (50\,Hz repetition rate) in the improved shielding geometry 
configuration for the mercury jet target. Also shown is the 
estimated power deposition within various regions (blue italics denote the power 
deposited in the beam pipe).
\label{fig:IDS120jPowerDep}}
\end{figure}
Table~\ref{tab:IDS120jPower} provides a detailed breakdown of the deposited power
in various subregions of the new geometry,
where the uncertainties are estimated by using different initial random number seeds
for the simulation. The first SC coil experiences a total power
of 0.4\,kW, with the other coils experiencing much lower values, giving a total
approximately equal to 0.6\,kW. These correspond to energy densities below 0.1\,mW/g, 
which permits at least a 10 year (of $2 \times 10^7$\,s each) operational lifetime 
against radiation damage to their organic (MgO) insulators, assuming the maximum 
permitted radiation dose is 10\,MGy~\cite{ref:Schultz}. 

About 2.4\,MW is deposited in the combined shielding and beam pipe sections. Removing the
heat load from these materials will prove quite challenging. The target and mercury
pool receives a total power deposition approximately equal to 0.9\,MW, which must
be dissipated in a heat exchanger in the mercury flow return loop.
The downstream Be-He-Be window (1.5\,cm thick) receives a large energy deposition of 9\,kW.
About 250\,kW will continue into the downstream pion--muon transport system. This will 
contain scattered, high energy protons that must be removed by a chicane and
absorber system in order to stop radiation damage to other 
accelerator components~\cite{ref:Chicane}.

Table~\ref{tab:IDS120jPower} also shows the deposited power in the various regions when
alternative materials are used for the target. In general, most of the
differences are seen in the energy deposition in the target itself, the shielding 
and the liquid/powder collection pool.
Less protons interact with the liquid gallium jet target, going directly into
the pool reservoir. This very slightly reduces the overall power deposited in the
shielding and beam pipe sections. The reduced
density in the tungsten powder jet also means that more energy is deposited
in the powder collection reservoir (assumed to have the same geometry
as the mercury pool), although not as much as was the case
for the gallium target. In contrast, the solid tungsten target  
itself experiences a much larger energy deposition,
corresponding to a quarter of the total beam power. Experimental work
using fast, high current pulses passing through tungsten wires has demonstrated that 
a 20\,cm-long, 2\,cm diameter tungsten target will be able to withstand
the peak stresses from such an energy deposition, allowing a target lifetime
of at least three years~\cite{ref:WSolidShock}. The removal
of the reservoir pool for the solid target means that about 5\% 
more energy is deposited in the shielding, with a slight
reduction in the radiation dose absorbed by the downstream beam pipe.

\section{Failure modes\label{sec:Failure}}

It is important to know how the 4\,MW total beam power will be distributed 
within the new target station geometry for specific (baseline) failure modes:
when the magnetic field fails, when the mercury jet stops flowing, and when both
happen together. Table~\ref{tab:IDS120jFailures} shows comparisons of the
average power deposition in various regions of the target station
between normal and failure-mode operating scenarios.

{
\begin{table*}[hbt!]
\caption{Power deposition in various regions of the increased shielding
geometry configuration for the mercury jet for a range of operational 
failure modes.
\label{tab:IDS120jFailures}}
\begin{ruledtabular}
\begin{tabular}{lllll}
& \multicolumn{4}{c}{Deposited power (kW)}\\
Region & Nominal & No Hg jet & No $\underline{B}$ & No $\underline{B}$ \& no Hg jet \\
\hline
SC coils 1--12 & $0.57 \pm 0.05$ & $0.43 \pm 0.05$ & $0.50 \pm 0.05$ & $0.51 \pm 0.05$ \\
Lower shielding SC 1--3 ($r < 50$\,cm, $z < 83$\,cm) & $1284.4 \pm 8.3$ & $42.0 \pm 0.3$ & $842.8 \pm 5.8$ & $842.9 \pm 5.5$ \\
Lower shielding SC 1--3 ($r < 50$\,cm, $z > 83$\,cm) & $234.2 \pm 3.5$ & $668.3 \pm 2.9$ & $1370.3 \pm 9.4$ & $1371.2 \pm 9.4$ \\
Upper shielding SC 1--3 ($r > 50$\,cm) & $58.3 \pm 0.8$ & $120.7 \pm 0.8$ & $61.6 \pm 0.8$ & $61.7 \pm 0.7$ \\
Shielding for SC 4--6 & $38.0 \pm 1.7$ & $15.0 \pm 1.0$ & $23.8 \pm 1.4$ & $23.9 \pm 1.3$ \\
Shielding for SC 7--9 & $11.0 \pm 0.7$ & $0.8 \pm 0.3$ & $3.1 \pm 0.5$  & $3.0 \pm 0.4$ \\
Shielding for SC 10--12 & $7.4 \pm 0.7$ & $0.2 \pm 0.1$ & $1.6 \pm 0.3$ & $1.6 \pm 0.4$ \\
Beam pipe up to $z = 0$\,cm & $352.3 \pm 2.9$ & $0.2 \pm 0.1$ & $0.7 \pm 0.1$ & $0.7 \pm 0.1$ \\
Beam pipe from $z = 0$\,cm to end of taper & $397.6 \pm 3.6$ & $218.8 \pm 1.7$ & $1013.5 \pm 5.0$ & $1012.8 \pm 5.5$\\
Beam pipe from end of taper & $21.7 \pm 0.9$ & $0.5 \pm 0.2$ & $4.6 \pm 0.4$ & $4.6 \pm 0.5$ \\
Lower shielding vessel for SC 1--3 ($r < 50$\,cm) & $7.6 \pm 0.2$ & $12.0 \pm 0.2$ & $8.0 \pm 0.2$ & $8.0 \pm 0.2$ \\
Upper shielding vessel for SC 1--3 ($r > 50$\,cm) & $6.0 \pm 0.1$ & $11.8 \pm 0.1$ & $6.3 \pm 0.1$ & $6.2 \pm 0.1$ \\
Shielding vessel for SC 4--6 & $3.5 \pm 0.3$ & $1.9 \pm 0.2$ & $2.6 \pm 0.3$ & $2.6 \pm 0.2$ \\
Shielding vessel for SC 7--9 & $0.8 \pm 0.1$ & $0.1 \pm 0.1$ & $0.2 \pm 0.1$ & $0.2 \pm 0.1$ \\
Shielding vessel for SC 10--12 & $0.5 \pm 0.1$ & $< 0.1$ & $0.1 \pm 0.1$ & $0.1 \pm 0.1$ \\
Pool reservoir container & $10.5 \pm 0.3$ & $10.8 \pm 0.1$ & $14.1 \pm 0.2$ & $14.2 \pm 0.2$ \\
Pool reservoir & $460.8 \pm 9.7$ & $2603.2 \pm 4.8$ & $235.9 \pm 2.5$ & $237.5 \pm 2.4$ \\
Jet/target & $416.8 \pm 2.4$ &  & $1.9 \pm 0.1$ &  \\
Be window & $8.9 \pm 0.1$ & $0.1 \pm 0.1$ & $0.3 \pm 0.1$ & $0.3 \pm 0.1$ \\
Total & $3320.7 \pm 14.4$ & $3707.0 \pm 6.0$ & $3591.9 \pm 12.5$ & $3591.9 \pm 12.6$ \\
\end{tabular}
\end{ruledtabular}
\end{table*}
}

Under normal conditions, noninteracting protons from the beam
have a trajectory that enters the mercury pool reservoir. This is illustrated
by the case when there is no mercury jet target, in which more than half of the total
beam power (2.6\,MW) is deposited in the pool. This will produce significant 
agitation of the pool surface with splashes of radial velocities 
expected to approach 50\,m\,s$^{-1}$~\cite{ref:HgSplash}.
In contrast, when there is no magnetic field present
to steer the proton beam (and any secondary
charged particles), the energy deposition for the combined mercury jet and pool
system decreases by roughly a factor of 3. The proton beam almost completely 
misses the mercury jet target and hits only part of the mercury pool, instead
dramatically increasing the energy deposition in the surrounding shielding
(1.6 to 2.3\,MW) and the nearby beam pipe section (from 0.4 to 1\,MW).
For the scenario when there is no mercury jet and no magnetic field, the results
are essentially identical to the case when only the magnetic field is turned off,
owing to the fact that the proton beam will miss the target in both cases.

\section{Radiation Safety\label{sec:RadDose}}

Any construction of the target station must take into account the
safety requirements of the surrounding environment to prevent radiation
contamination of the soil (and ground water). The entire target station must
be enclosed within a concrete shielding structure of adequate thickness
to stop radiation from escaping; the effective total radiation dose must 
not exceed 1\,mSv per year~\cite{ref:CERNSafety}, which is equivalent 
to a continual residual dose of 0.1\,$\mu$Sv/hr.

Figure~\ref{fig:IDS120jRadDose} shows the estimated total radiation ambient dose 
equivalent from the 4\,MW target station after a total irradiation time of 
$2\times10^{7}$\,s (1 year), using the FLUKA simulation code and 
AMB74 conversion factors~\cite{ref:AMB74}.
The target station is surrounded by a concrete shielding structure (tunnel) that
is modeled as three connecting cylindrical sections, extending from $z=-8$\,m up to 
$z=30$\,m, with an outer radius of 6\,m. This tunnel creates a barrier
between the surrounding rock, assumed to be molasse soil based at the CERN site, and the
radiation generated by the target system. The chemical composition (with mass fractions
in parenthesis) assumed for the concrete is O (51.1\%), Si (35.8\%), Ca (8.6\%),
Al (2.0\%), Fe (1.2\%), H (0.6\%), C (0.4\%) and Na (0.3\%), while the composition of the 
molasse soil is taken to be O (49.2\%), Si (19.8\%), Ca (9.3\%), Al (6.4\%),
C (4.9\%), Fe (4.1\%), Mg (3.5\%), K (1.9\%), Na (0.6\%), Mn (0.2\%) and 
Ti (0.1\%)~\cite{ref:Molasse}. The densities assumed for the concrete 
and molasse are 2.35\,g/cc and 2.4\,g/cc, respectively.
For a range of cooling decay times, it can be seen that
no radiation escapes the concrete shielding, which means that there 
will be minimal activation of the surrounding soil and ground water.
We can also infer that remote handling will be mandatory
for maintaining the target system; even at a radial distance of 2\,m from
the interaction region, the residual dose rate after 1 year of cooling is of the order of 
10\,mSv/hr, which greatly exceeds the safety limit for radiation workers 
(20\,mSv/yr)~\cite{ref:CERNSafety}.

\begin{figure*}[!hbt]
\includegraphics[width=\columnwidth]{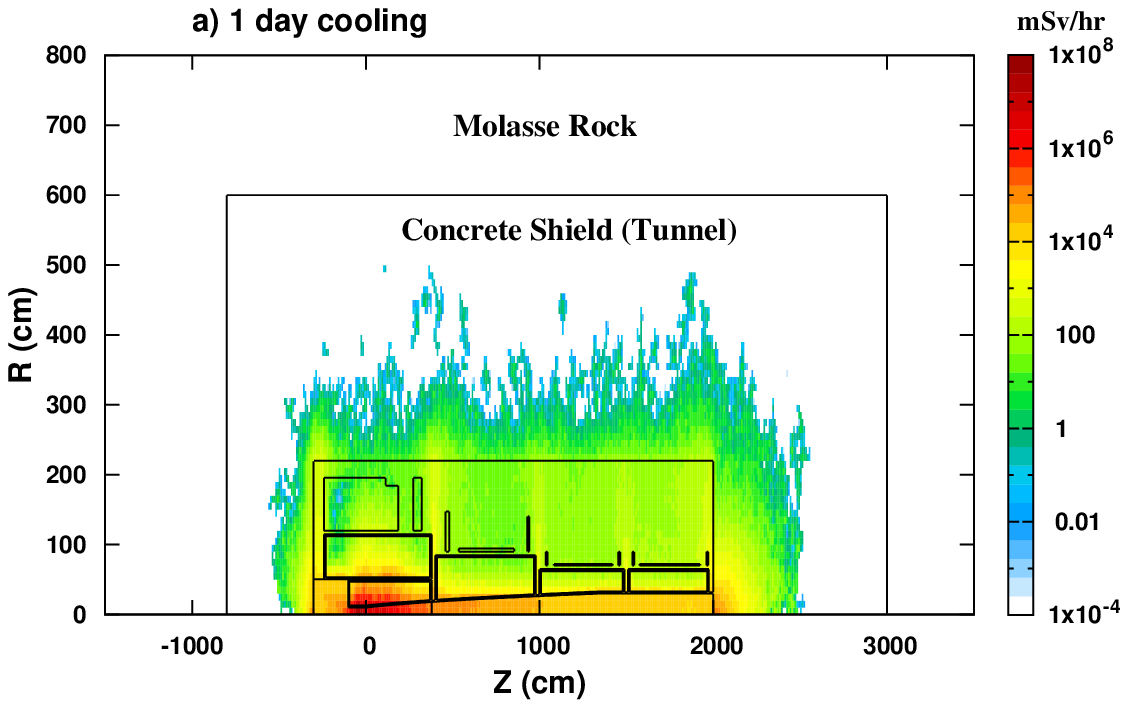}
\includegraphics[width=\columnwidth]{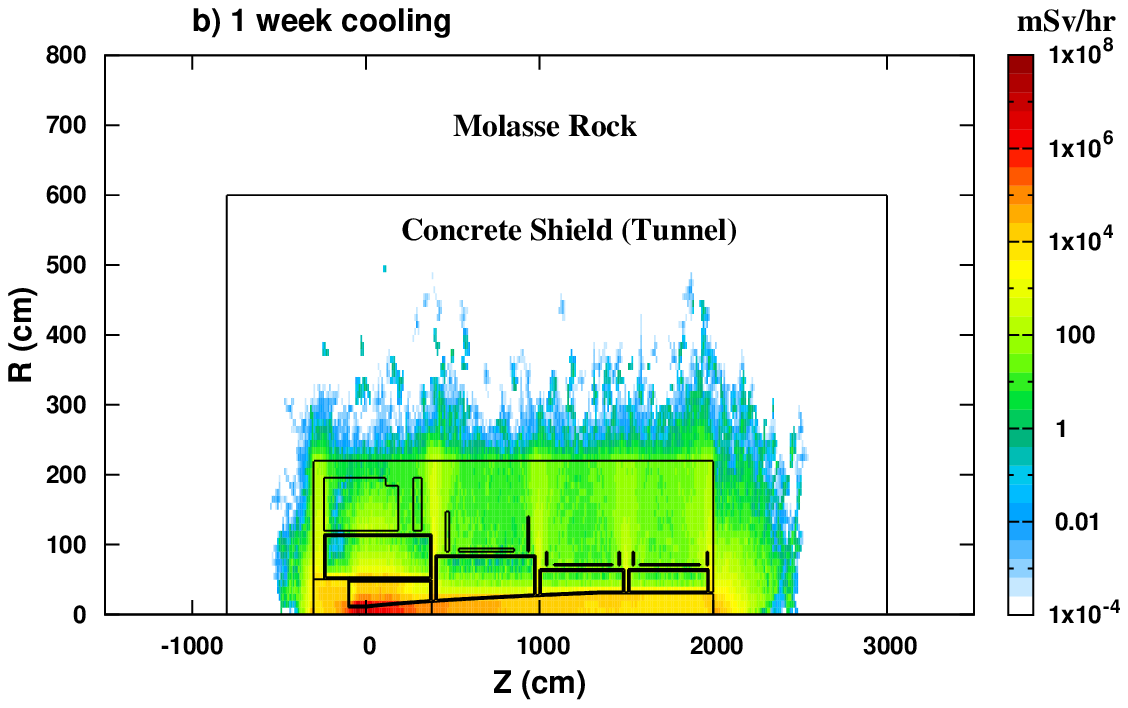}
\includegraphics[width=\columnwidth]{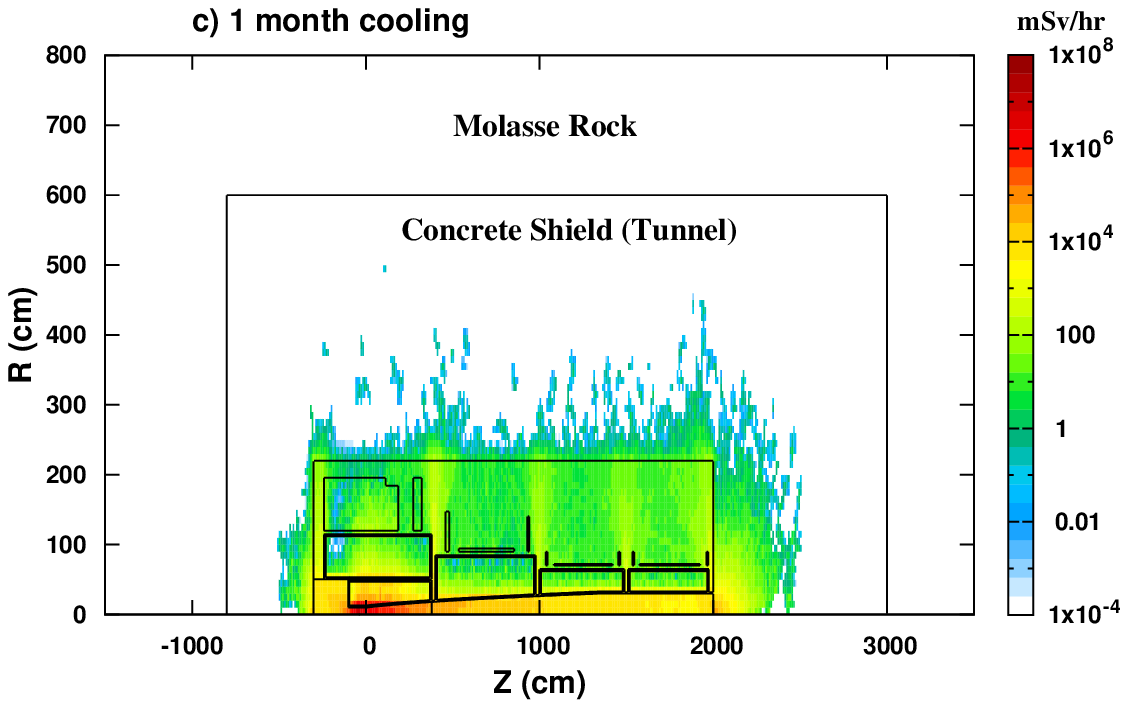}
\includegraphics[width=\columnwidth]{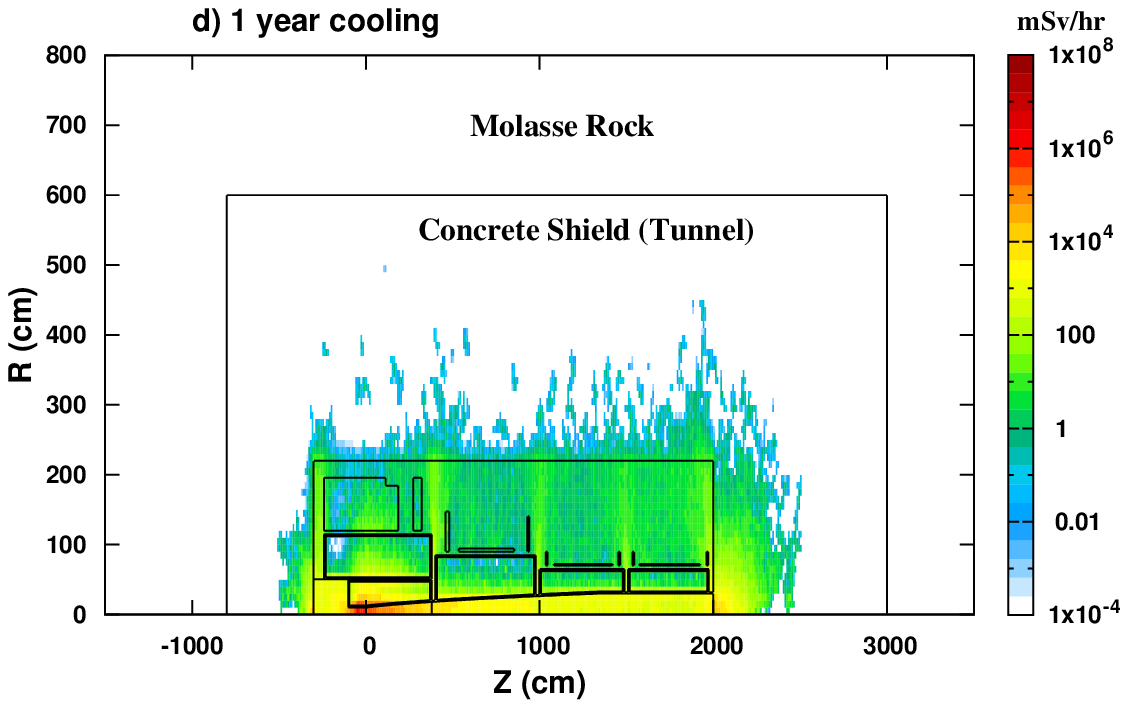}
\caption{Ambient dose equivalent rates for the target station, concrete
shielding tunnel, and surrounding underground rock.
\label{fig:IDS120jRadDose}}
\end{figure*}

\section{Summary}

We have presented FLUKA and MARS simulation studies of the pion production and 
energy deposition in the Neutrino Factory target station. Compared
to the original Study 2 geometry, a doubling of the lateral shielding thickness 
is needed to protect the superconducting coils from radiation.
In addition, a reduction in the focusing magnetic field from 20\,T down to 15\,T does
not affect the pion production efficiency, provided the inner bore radius is increased.
Alternative target materials were also investigated, such as liquid gallium, and powdered
and solid tungsten. Each of these offer comparable muon yields, with some
differences observed in the power deposited in the target, collection pool, and 
surrounding shielding. We have investigated what effect various operational
failure modes have on the deposited power in the target station.
Finally, we have provided estimates of the amount of concrete
shielding that will be needed to protect the environment from the high radiation 
generated by the target station, with remote handling mandatory for any
maintenance work.

\begin{acknowledgments}
We acknowledge the financial support of the European Community under
the European Commission Framework Programme 7 Design Study: EUROnu,
Project No 212372. We also
thank colleagues from the International Design Study (IDS-NF)
collaboration for fruitful discussions concerning this work.
\end{acknowledgments}

\end{document}